\documentclass[pra,twocolumn,nofootinbib,preprintnumbers,superscriptaddress]{revtex4-1}

\usepackage{natbib}
\usepackage{amssymb,amsbsy,amsmath,amsfonts}
\usepackage{mathrsfs}
\usepackage{graphicx}
\usepackage{epsf,epsfig,float,latexsym,amsthm,fancyhdr,rotating}
\usepackage{graphics,psfrag,longtable}
\usepackage{slashed}
\usepackage{esint}
\usepackage{nicefrac}
\usepackage{braket}
\usepackage{mathtools}

\newcommand{\nn}{\nonumber}

\newcommand{\be}{\begin{equation}}
\newcommand{\ee}{\end{equation}}

\usepackage[colorlinks,citecolor=blue,linktoc=all,linkcolor=cyan]{hyperref}

\begin{document}
\title {Delta baryon photoproduction with twisted photons }

\author{Andrei Afanasev}

\affiliation{Department of Physics,
The George Washington University, Washington, DC 20052, USA}

\author{Carl E. Carlson}

\affiliation{Physics Department, William \& Mary, Williamsburg, Virginia 23187, USA}

\begin{abstract}
A future gamma factory at CERN or accelerator-based gamma sources elsewhere can include the possibility of energetic twisted photons, which are photons with a structured wave front that can allow a pre-defined large angular momentum along the beam direction.  Twisted photons are potentially a new tool in hadronic physics, and we consider here one possibility, namely the photoproduction of $\Delta$(1232) baryons using twisted photons.  We show that particular polarization amplitudes isolate the smaller partial wave amplitudes and they are measurable without interference from the terms that are otherwise dominant.
\end{abstract}
\date{\today
}
\maketitle


\section{Introduction}


Twisted photons are examples of light with a structured wave front that can produce transitions with quantum number changes that are impossible with plane wave photons.  One can envision a number of applications in the field of hadron structure that would require sources of twisted photons with MeV-GeV energy scales. Such energies are achievable via Compton up-conversion in high-energy electron collisions with twisted optical photons \cite{Jentschura:2011ih,Petrillo16} or in twisted-photon collisions with high-energy ions, as recently suggested for CERN Gamma Factory \cite{Budker20}. Presently, HIGS facility is making important steps toward twisted-photon generation \cite{liu2020orbital}, opening new opportunities for nuclear physics studies. 

This article will focus on how twisted light may contribute to measuring small but important contributions to the electromagnetic production of $\Delta$(1232) baryons from nucleon targets.
 
More specifically, twisted photons are states with total angular momentum whose projection along the direction of motion can be any integer, $m_\gamma$, times $\hbar$.  The Poynting vector or momentum density of these states swirls about a vortex line, and the intensity of the wavefront is typically zero or very small on the vortex line.  (Indeed, the ``hole'' in the middle of the wavefront can find applications seemingly unconnected to the swirling of the states, as in stimulated emission spectroscopy studies; see~\cite{2021arXiv210407095D}.)  

In photoabsorption, the photon's projected angular momentum $m_\gamma$ is transferred to the target system and may be shared between internal excitation of the final state and orbital angular momentum of the final state's overall center-of-mass.  That the final internal angular momentum projection can differ by $m_\gamma$ units from that of the target allows transitions quite different from the plane wave case, where the photon necessarily transfers $m_\gamma = \Lambda = \pm 1$ to the final excitation. Here $\Lambda$ is the helicity or spin along the direction of motion of the photon and that $\Lambda = \pm 1$ only is a standard fact for plane waves.

That large projected angular momentum transfer works in practice for atoms has been shown experimentally and reported in~\cite{2016NatCo...712998S}, where optical orbital angular momentum was transferred to bound electrons, exciting $S_{1/2}$ to $D_{5/2}$ transitions in $^{40}$Ca ions with quantum number changes beyond what a plane wave photon could induce.  Further, the sharing of final state angular momentum between internal and overall degrees of freedom, when the ion was offset from the vortex line, was also measured and matched well with theoretical studies~\cite{Afanasev_2018,Solyanik-Gorgone_2019}.  

\begin{figure}[h]
\begin{center}

\includegraphics[width = 0.7 \columnwidth]{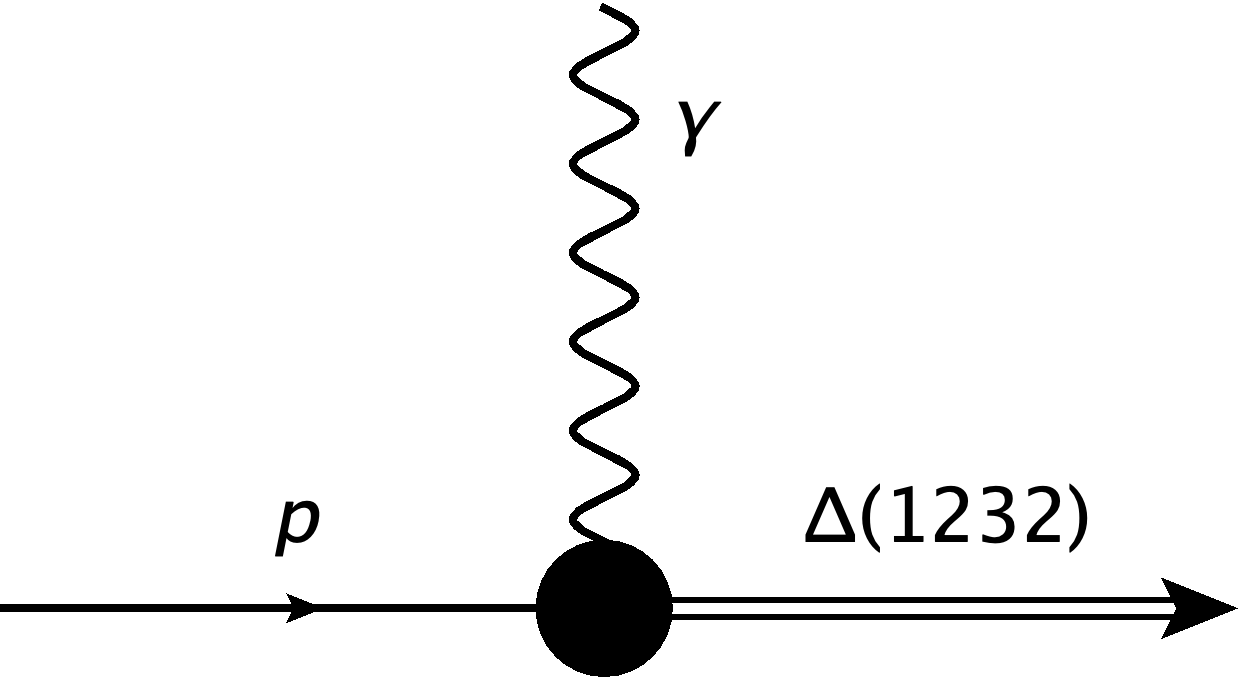}
\caption{Photoexcitation of $\Delta(1232)$ baryon on a proton target. }
\label{fig:diagram}
\end{center}
\end{figure}

The $\Delta(1232)$ is a spin 3/2 baryon that can be photoexcited from the nucleon (Fig.\ref{fig:diagram}) via $M1$ (with related notations $M_{1+}$ and $G_M^*$) and $E2$ (ditto with $E_{1+}$ or $G_E^*$) transitions.  For a thorough review of electromagnetic excitation of the $\Delta(1232)$, see~\cite{Pascalutsa:2006up}.   In simple models, the $\Delta$ is dominantly a spatial $S$-state with a spin-3/2 spin wave function.  This can be obtained from the nucleon by a simple spin flip, which an $M1$ transition can do, and there is a large N to $\Delta$ $M1$ amplitude.  The $E2$ transition requires two units of orbital angular momentum, and must involve the small $D$-wave spatial component of the $\Delta$ or of the nucleon.  Accurate knowledge of the $E2$ size would help elucidate the $\Delta$ composition and hadron structure generally.  

Currently the particle data group~\cite{Zyla:2020zbs} quotes an $G_E^*/G_M^*$ ratio in the 2-3 \% range, based on plane wave photon cross section measurements where the $E2$ is neither the only nor the dominant contribution.  We shall show that twisted photons can in principle produce signals there the $E2$ is the only contributor.

In much of what follows, it is more natural to describe plane wave $\Delta$ photoproduction using helicity amplitudes $\mathcal M^\text{(pw)}_{m_i \Lambda}$, where $\Lambda$ is the helicity of the photon and $m_i$ is the spin of the nucleon target along the photon's momentum direction.  There are two independent helicity amplitudes
\begin{align}
\mathcal M^\text{(pw)}_{1/2,1}   &\, \propto \sqrt{3} \left( G_M^* + G_E^* \right), \nn\\
\mathcal M^\text{(pw)}_{-1/2,1} &\, \propto  \   G_M^* - 3 G_E^*  	,
\end{align}
and others can be obtained by parity transformation
\be
\mathcal M^\text{(pw)}_{-m_i,  -\Lambda} = - \mathcal M^\text{(pw)}_{m_i \Lambda}	\,.
\ee
In this notation,  our goal is to use twisted photons to isolate physically achievable situations where the helicity amplitudes combine with the $M1$ contributions canceling and the $E2$ not.

The calculation of the N to $\Delta$ transition with twisted photons is in many ways analogous to atomic calculations.  However, the most straightforward atomic calculations are made in a no-recoil limit, which gives accurate results for targets that are quite massive compared to the photon energy.   For the the N to $\Delta$ transitions we do take account of the recoil,  eventually finding that the recoil corrections are small, at the level of a few \%.  Also, for the apparently simplest atomic analog, an $S_{1/2}$ to $D_{3/2}$ transition in a single electron atom, the $M1$ does not contribute at all to leading order.  However, our crucial results are based on the properties of the twisted photon and on the quantum numbers and rotation properties of the hadronic states.  A more exactly analogous atomic analog can be found among multi-electron atoms.  In particular, there is an Boron-like example where the calculated $M1$ and $E2$ amplitudes are about the same size~\cite{rynkun2012energies}, and we have elsewhere shown how twisted photons could be instrumental in measuring these amplitudes~\cite{2018PhRvA..97b3422A}.  

In the following, Sec.~\ref{sec:twist} will contain some background material on twisted photons, which may be skipped or skimmed by readers already expert.  Sec.~\ref{sec:wo} will, for the sake of beginning simply, find results for twisted photon induced N to $\Delta$ transitions in the no-recoil limit.  Sec.~\ref{sec:with} will include proper relativistic consideration of the $\Delta$ recoil, and will show that the corrections are visible on some plots but that they are not large.
Some final comments will appear in Sec.~\ref{sec:end}.


\section{Twisted photons}		\label{sec:twist}


A twisted photon is a state whose wavefront travels in a definite direction and which has arbitrary integer angular momentum along its direction of motion.  Reviews may be found, for example, in~\cite{2011AdOP....3..161Y,Bliokh:2015doa}.  In addition, the states studied are usually monochromatic, and at a theoretical level, one can choose between Laguerre-Gaussian or Bessel versions of these states.  We here use the latter;  experience has shown that results are numerically nearly the same either way, whereas analytic expressions are simpler for Bessel modes.  

\begin{figure}[t]
\begin{center}

\includegraphics[width = 0.98 \columnwidth]{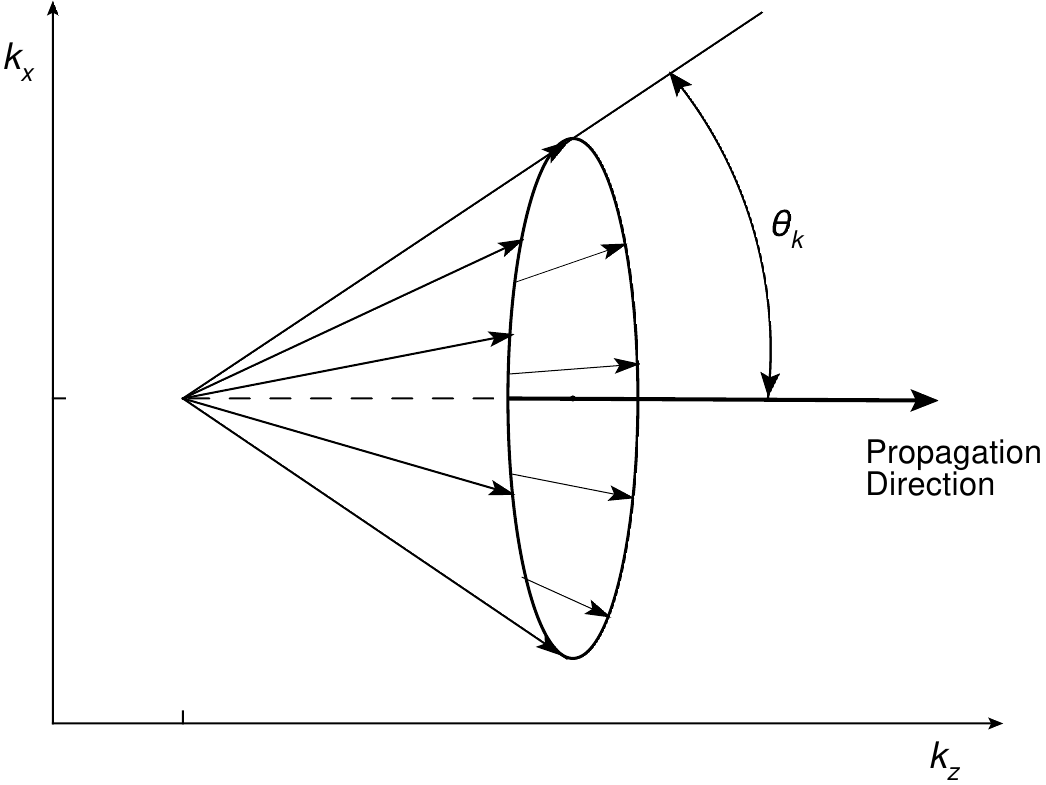}
\caption{A twisted photon state in wavenumber space or momentum space.}
\label{fig:tp}
\end{center}
\end{figure}

Bessel photons, in addition to the twistedness and monochromaticity, are nondiffracting exact solutions to the Helmholtz equation.   The states can be most simply written in wave number space, or momentum space, where they can be represented as a collection of plane wave photons, each with the same value of $k_z$, where $z$ is the direction of propagation of the state,  each with the same magnitude of transverse momentum $| \vec k_\perp | = \kappa $, and hence each with the same polar angle or pitch angle $\theta_k = \arctan(\kappa/k_z)$, but differing azimuthal angles $\phi_k$.   The set of wave vectors thus form a right circular cone in momentum space, Fig.~\ref{fig:tp}.

The state is~\cite{Jentschura:2010ap,Jentschura:2011ih,Afanasev:2013kaa}
\begin{align}
\label{eq:twisteddefinition}
| \kappa m_\gamma k_z \Lambda \vec b \rangle 
&= A_0  \int \frac{d\phi_k}{2\pi} (-i)^{m_\gamma} e^{im_\gamma\phi_k - i \vec k \cdot \vec b}  \,
	|\vec k, \Lambda\rangle	\,,
\end{align}
where $m_\gamma$ is the total angular momentum in the $z$-direction,  $| \vec k, \Lambda \rangle $ is a plane wave photon state with
\be
\braket{  \vec k', \Lambda' | \vec k, \Lambda } = (2\pi)^3 \, 2 \omega \delta_{\Lambda' \Lambda}
	\delta^3(\vec k' - \vec k)	\,,
\ee
$\Lambda$ is the helicity of each component state,  $\omega$ is the angular frequency of the monochromatic state, and $A_0$ is a normalization chosen, for example, in~\cite{Jentschura:2010ap,Jentschura:2011ih} as $A_0 = \sqrt{\kappa/(2\pi)}$.  The state written has a vortex line passing through the point $\vec b = (b_x, b_y, 0 )$, where we might instead give a magnitude 
$b$ and azimuthal angle $\phi_b$.

With this state normalization, the electromagnetic potential for a plane wave photon is
\be
\braket{ 0 |  A_\mu(x) | \vec k, \Lambda } = \epsilon_\mu(\vec k, \Lambda) e^{-i k x},
\ee
where $\epsilon_\mu(\vec k, \Lambda)$ is a unit polarization vector for a photon of the stated momentum and helicity.   The vector potential for the twisted state in coordinate space can now be worked out, and in cylindrical coordinates and for $\vec b = 0$  the components are
\begin{align}
\label{eq:twA}
 A_\rho &= i \frac{ A_0 }{\sqrt{2}}	\,
	e^{i(k_z z - \omega t + m_\gamma \phi )}		\nn\\
&	\times
	\left[  \cos^2 \frac{\theta_k}{2}  J_{m_\gamma-\Lambda}(\kappa \rho)
	+ \sin^2 \frac{\theta_k}{2}  J_{m_\gamma + \Lambda}(\kappa \rho)	\right] ,    \nonumber\\
 A_\phi &= - \Lambda \, \frac{ A_0 }{\sqrt{2}}	\,
	e^{i(k_z z - \omega t + m_\gamma \phi )}		\nn\\
&	\times
	\left[  \cos^2 \frac{\theta_k}{2}  J_{m_\gamma-\Lambda}(\kappa \rho)
	- \sin^2 \frac{\theta_k}{2}  J_{m_\gamma + \Lambda}(\kappa \rho)	\right] ,    \nonumber\\
 A_z &= \Lambda \, \frac{ A_0 }{\sqrt{2}}	\,
	e^{i(k_z z - \omega t + m_\gamma \phi )}	
	\sin\theta_k  \,  J_{m_\gamma}(\kappa \rho) ,
\end{align}
with $\vec E = i \omega \vec A$ and $\vec B = - i \Lambda \vec E$.   The $J_\nu$ are Bessel functions.  For general impact parameter, let $\vec \rho$ be the transverse coordinates and let $\rho \to | \vec \rho - \vec b |$.

One sometimes makes a paraxial approximation, wherein the pitch angle $\theta_k$ is small, and one keeps only the leading nontrivial terms in $\theta_k$.  We will usually not do this.  Additionally, one can find exact twisted wave solutions to the Helmholtz equation where the smaller terms (for small $\theta_k$) in $A_\rho$ and $A_\phi$ above are absent~\cite{2019PhRvA..99b3845Q}, with the expenses of changing the longitudinal component and of not having all the component photons in momentum space have the same helicity.  We will find the uniformity in helicity useful, and so stay with the forms above.  

The time averaged Poynting vector is
\begin{align}
\braket{S_\rho} &= 0 ,    \nonumber\\
\braket{S_\phi} &= \frac{ \omega^2 A_0^{\,2} }{2}  \,  \sin\theta_k J_{m_\gamma}(\kappa \rho)  \nn\\
&	\times
	\left[  \cos^2 \frac{\theta_k}{2}  J_{m_\gamma-\Lambda}(\kappa \rho)
	+ \sin^2 \frac{\theta_k}{2}  J_{m_\gamma + \Lambda}(\kappa \rho)	\right] ,    \\ \nonumber
\braket{S_z} &= \frac{ \omega^2 A_0^{\,2} }{2}
	\left[  \cos^4 \frac{\theta_k}{2}  J^2_{m_\gamma-\Lambda}(\kappa \rho)
	- \sin^4 \frac{\theta_k}{2}  J^2_{m_\gamma + \Lambda}(\kappa \rho)	\right].
\end{align}

To close this section and prepare for the next, we work out the photoabsorption amplitude involving the twisted photon in the limit where the $\Delta$ is treated as very heavy and its recoil velocity is neglected.  The necessary manipulations mirror the atomic physics case worked out in~\cite{scholz2014absorption}.  We wish to obtain the amplitude
\be
\mathcal M = 
\braket{ \Delta(m_f) \,  | \mathcal H(0) | \, N(m_i); \, \gamma( \kappa m_\gamma k_z \Lambda \vec b) },
\ee
where $\mathcal H$ is the interaction Hamiltonian density.  The nucleon is at rest, and its spin projection along the $z$-axis is labeled as $m_i$.  Similarly, we label the spin projection of the $\Delta$ along the same axis as $m_f$.  The twisted photon state can be expanded in plane waves, as in Eq.~\eqref{eq:twisteddefinition}, and the plane wave photon states can be obtained by rotations of states with momenta in the $z$-direction,
\be
\ket{ \vec k, \Lambda } = R(\phi_k,\theta_k,0)  \ket{  k \hat z, \Lambda }	,
\ee
which follows~\cite{Jentschura:2010ap,Jentschura:2011ih} in using the Wick phase conventions of 1962~\cite{Wick:1962zz}.

The Hamiltonian is rotation invariant. Rotations of the nucleon states are given in terms of the Wigner functions,
\be
R^\dagger(\phi_k,\theta_k,0) \ket{ N(m_i) } = e^{i m_i \phi_k} 
	\sum_{m'_i}  d^{1/2}_{m_i,m'_i}(\theta_k) \ket{ N(m'_i) }	,
\ee
and rotations of the $\Delta$ can be given analogously---if the recoil velocity of the $\Delta$ is neglected.  One obtains
\begin{align}
\label{eq:nramplitude}
\mathcal M &= A_0 (-i)^{m_f-m_i}  e^{ i (m_\gamma + m_i - m_f) \phi_b }   
	J_{m_f -m_i -m_\gamma}(\kappa b)
		\nn\\[1 ex]
&\quad	\times \sum_{m'_i}   d^{3/2}_{m_f ,m'_i+\Lambda} (\theta_k)	\, d^{1/2}_{m_i,m'_i}(\theta_k)  \,
	\mathcal M^{\text{(pw)}}_{m'_i,\Lambda}	.
\end{align}
The plane wave amplitude is defined from
\be
\braket{ \Delta(m'_f) \,  | \mathcal H(0) | \, N(m'_i); \, \gamma( k \hat z, \Lambda) }
	= \mathcal M^{\text{(pw)}}_{m'_i,\Lambda}	 \,	\delta_{m'_f, m'_i + \Lambda}	.
\ee
The $\delta$-function follows because all the spins and momenta in the plane wave amplitude are along the $z$-direction.

In terms of the Jones-Scadron form factors~\cite{Jones:1972ky,Pascalutsa:2006up},
\begin{align}
\label{eq:pwamplitudes}
    \mathcal M^\text{(pw)}_{1/2,1} &= 
        - \frac{3 e E_\gamma}{2}  \sqrt{\frac{2}{3}}
            \left( G_M^* + G_E^* \right)  ,  \nn\\
    \mathcal M^\text{(pw)}_{-1/2,1} &= 
        - \frac{ \sqrt{3}\, e E_\gamma}{2} \sqrt{\frac{2}{3}}
            \left( G_M^* - 3 G_E^* \right)  , 
\end{align}
where an isospin Clebsch-Gordan factor $\sqrt{2/3}$ for the $p \to \Delta^+$ transition is included, and 
\be
E_\gamma = \frac{ m_\Delta^2 - m_N^2 }{ 2 m_N } = \frac{2\pi}{\lambda}
\ee
is the photon energy to excite a $\Delta$ from a nucleon at rest~\cite{jonesscadronnote}.


\section{Twisted amplitudes for $\Delta$ photoexcitation}		\label{sec:wo}


When a stationary proton absorbs a photon of the correct energy to produce a $\Delta(1232)$, the $\Delta$ recoils with the momentum acquired from the photon, at a not so negligible speed of about $0.27 c$.  None-the-less, we will start by calculating the twisted photon photoexcitation amplitudes neglecting the recoil.  We will find certain aspects of the no-recoil results are can be understood qualitatively, and others aspects can in simple ways be worked out analytically. In particular, that only the E2 amplitude can contribute to transitions with $\Delta m = m_f - m_i = \pm 2$ is easy to understand.  Then in the next section, we will show that for crucial observables, the recoil corrections change the result in fractional terms by only about the square of half the recoil speed (in units of $c$). 

The no-recoil result for the amplitude is like the result in an atomic transition.  It is given in terms of two independent plane wave amplitudes, a pair of Wigner functions, and a Bessel function, and is shown as Eq.~\eqref{eq:nramplitude} in the previous section.

We have prepared plots for a variety of $m_\gamma$, $\Lambda$, and $m_f = m_\Delta$, all for $m_i = m_p = -1/2$ and placed the results in a $4 \times 6$ grid in the Appendix. (Plots for $m_i = 1/2$ are identical if one makes the appropriate parity changes on the other quantum numbers.)

\begin{figure}[t]
\begin{center}
{\sf (a)} \hfill \,   

\includegraphics[width = 0.75\columnwidth]      {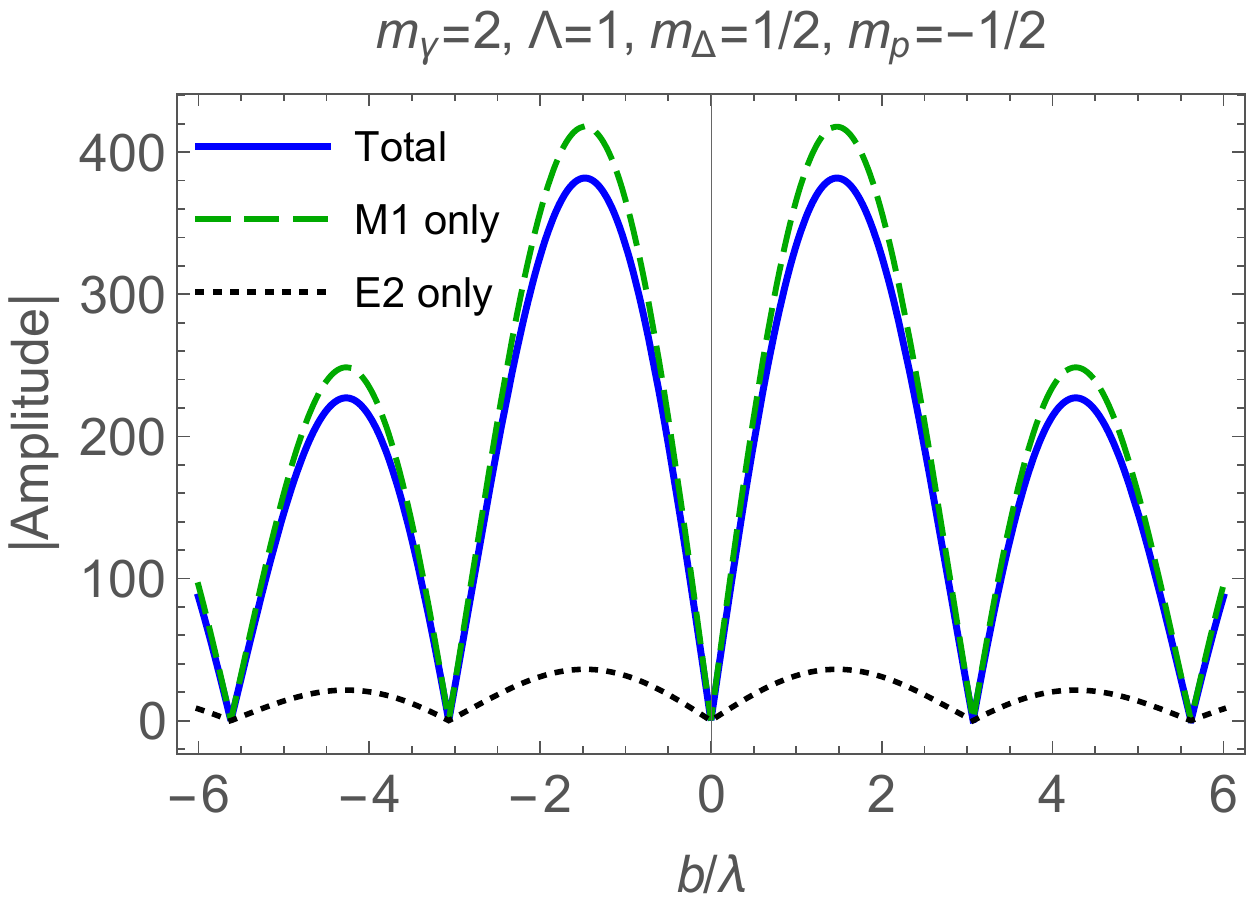}

{\sf (b)} \hfill \,

\includegraphics[width = 0.75\columnwidth]      {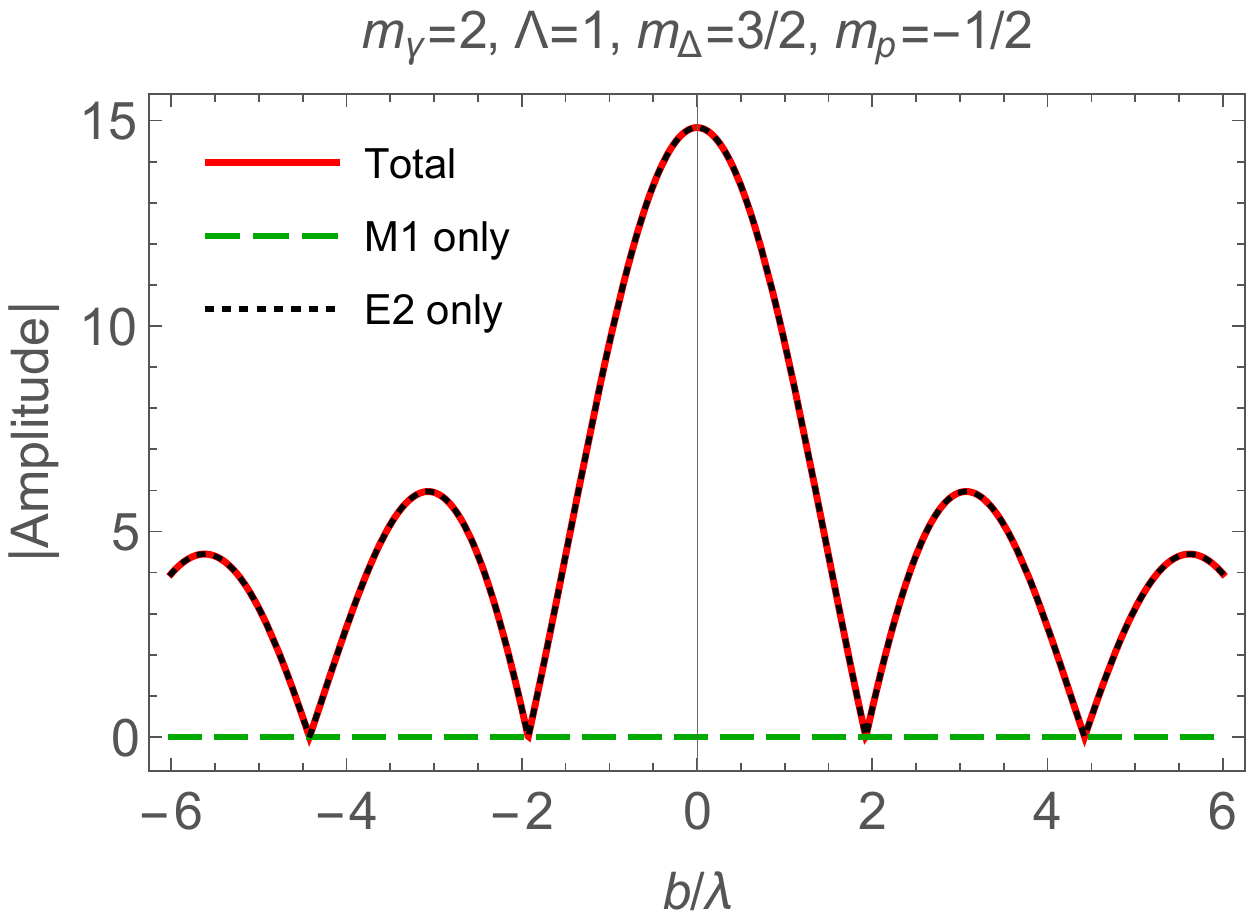}

{\sf (c)} \hfill \,

\includegraphics[width = 0.75\columnwidth]      {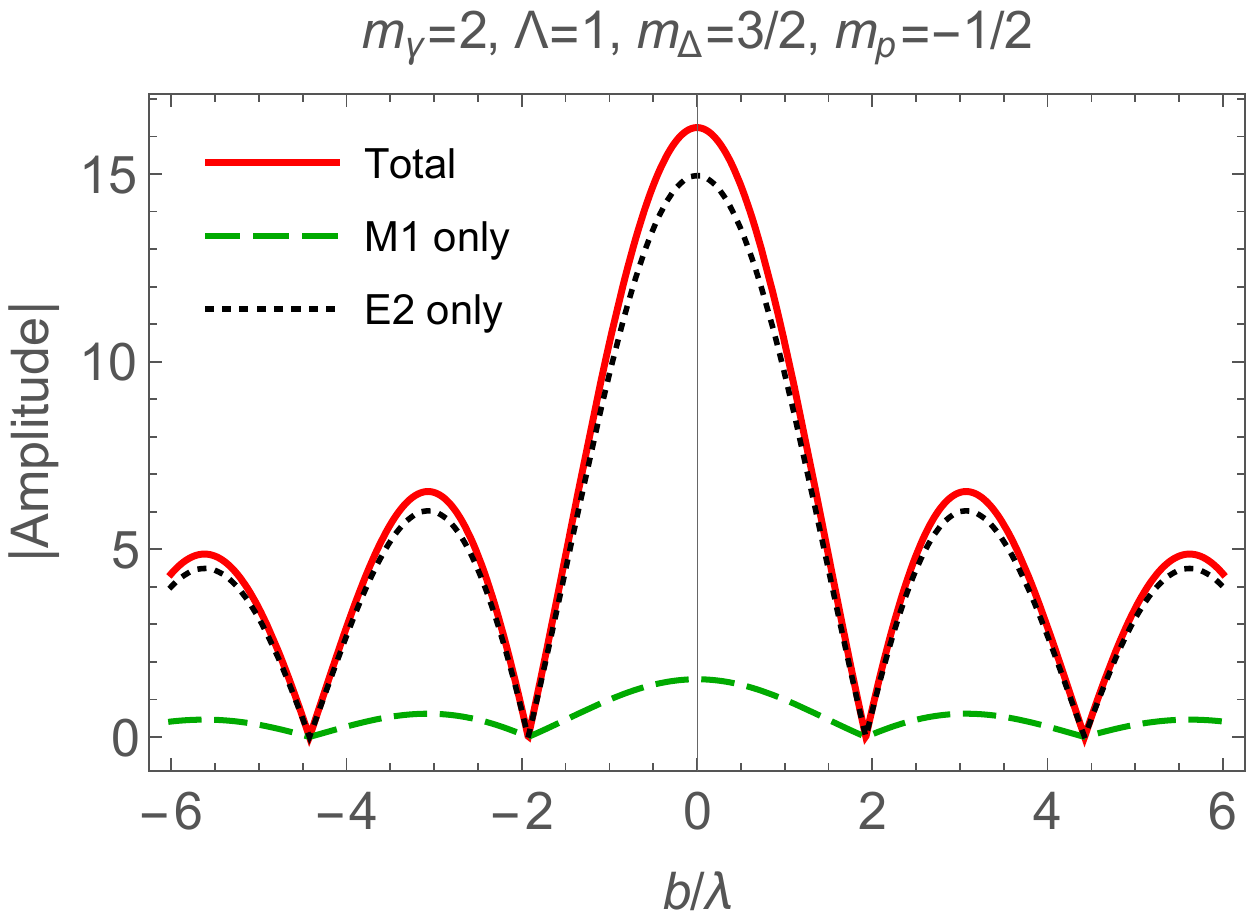}
\caption{Selected amplitude magnitude plots for twisted photon 
$+$ proton $\to \Delta$ vs.~displacement of the proton from the photon's vortex line, along (say) the positive and negative $x$-axis, in units of the photon wavelength ($3.65$ fm).  The amplitude units are arbitrary but are the same for each figure.  Each of these plots has $m_\gamma = 2$, $\Lambda = 1$, and $m_i = m_p = -1/2$.  Part (a) has $m_f=m_\Delta=1/2$ and parts (b) and (c) have $m_f=m_\Delta=3/2$.  (a) and (b) omit recoil corrections, as in Sec.~\ref{sec:wo}.  (c) includes recoil corrections, as in Sec.~\ref{sec:with}. The E2 only curve overlays or nearly overlays the Total in parts (b) and (c). The recoil corrections are visible but small.} 
\label{fig:bigamp}
\end{center}
\end{figure}

To understand the plots, we focus on two of them, shown in Fig.~\ref{fig:bigamp}(a) and (b).  The ordinate for both is the magnitude of the amplitude, the abscissa is the displacement of the vortex line of the twisted beam from the target proton, sweeping from the left along (say) the $x$-axis to the right in units of photon wavelength, $\lambda$.  Both of plots are for $m_\gamma = 2$, $\Lambda = 1$, and $m_i = -1/2$.  Fig.~\ref{fig:bigamp}(a) has $m_f = 1/2$ and Fig.~\ref{fig:bigamp}(b) has $m_f = 3/2$.  The plots are for pitch angle $\theta_k = 0.2$.  
 
General observations valid for all these plots, and not specific to the $N \to \Delta$, include

\noindent $\bullet$ \textit{on-axis observation:} If the target sits  on the vortex line, $b=0$, the angular momentum of the photon must all go into the internal excitation of the final state, i.e., $\Delta m = m_f - m_i = m_\gamma$.  If this cannot happen, the amplitude must be zero in the center, as in Fig.~\ref{fig:bigamp}(a) but not Fig.~\ref{fig:bigamp}(b).

\noindent $\bullet$ \textit{off-axis observation 1:} off-axis, the photon's total angular momentum can be shared by the internal excitation of the final state and by angular momentum of the final state overall center of mass~\cite{2014JOSAB..31.2721A}, and the amplitudes in general are not zero for $b \ne 0$.

\noindent $\bullet$ \textit{off-axis observation 2:} When the target is away from the vortex line, the target sees only a piece of the swirl, which can look  more unidirectional, like a plane wave.  The selection rules reflect this and transitions possible with plane wave quantum numbers tend to have larger peak amplitudes than other transitions.  This can be seen by comparing the vertical scales in Figs.~\ref{fig:bigamp}(a) and (b).

Specific to the $N \to \Delta$ transition are the separate contributions of the $M1$ and $E2$ amplitudes, which are shown in the plots.  Most interesting is Fig~\ref{fig:bigamp}(b), where the $M1$ does not contribute at all.  A measurement with the final $\Delta$ constrained to be in the projection $m_f = 3/2$ state would give a direct measurement of the $E2$ amplitude.

The plots are for $G_E^*/G_M^* = 3\%$, and Fig.~\ref{fig:bigamp}(a) shows the a more usual case where the $M1$ gives the major share of the amplitude.

Further regarding the absence of $M1$ contribution to the $\Delta m = m_f - m_i =2$ transition, there are only two terms in the sum for the amplitude, Eq.~\eqref{eq:nramplitude}, so it is easy to insert explicit Wigner functions and plane wave amplitudes in terms of $M1$ and $E2$, Eq.~\eqref{eq:pwamplitudes}, and demonstrate analytically that the $M1$ contribution cancels.  This is true for any $\Delta m = 2$ transition, and one can see this in the plots in the Appendix, where every plot in the right hand column has $\Delta m = 2$ and no $M1$ contribution.


\section{Recoil momentum corrections}		\label{sec:with}


The $\Delta$ recoils when it is produced by the proton absorbing the photon.  The rotations of the moving state are not given so simply as for a state at rest.  It will turn out that the corrections are not numerically large, but we want to consider them.

We find it helpful to consider the wave function of the initial nucleon state.  Even for situations that ultimately involve only plane waves, the manipulations of scattering theory are justified by using wave packets for the states---see for example~\cite{Taylor,Peskin:1995ev}---and then taking limits.  Here will use the twisted photon state as already given, but will more carefully consider the wave function of the proton state.  We want to localize it at the origin, but realize that there is a playoff between localization and the uncertainty of the target momentum.  We shall consider that the uncertainties in both position and momentum are small judged by to scales over which other quantities in the calculation vary~\cite{Taylor,Peskin:1995ev}.  

The amplitude is
\begin{align}
&\mathcal M =
\int (d^3 p) \widetilde\phi_i(\vec p_i)  \nn\\
&\hskip 3 em   \times
    \braket{\Delta(p_f,m'_f) | H | N(\vec p_i,m_i), 
        \gamma(\kappa m_\gamma k_z \Lambda \vec b ) }
\end{align}
where $H$ is the interaction Hamiltonian.  Inserting the expansion for the twisted photon gives
\begin{align}
\mathcal M &= A_0 \int \frac{d\phi_k}{2\pi} (-i)^{m_\gamma}
    e^{im_\gamma \phi_k - i \vec k \cdot \vec b} \,
    \widetilde\phi_i(\vec p_i)  \nn\\
&\hskip 2 em    \times 
    \braket{\Delta(p_f,m'_f) | \mathcal H(0) | N(\vec p_i,m_i),     \gamma(\vec k,\Lambda) }
\end{align}
where $\vec p_f = \vec p_i + \vec k$.  Further,
\begin{align}
\mathcal M &= A_0 \int \frac{d\phi_k}{2\pi} (-i)^{m_\gamma}
    e^{i(m_\gamma + m_i) \phi_k - i \vec k \cdot \vec b} \,
    \widetilde\phi_i(\vec p_f - \vec k)  \nn\\
&\hskip 2 em \times
    \mathcal M^\text{(pw)}_{m'_i \Lambda} \, 
    d^{1/2}_{m_i m'_i}(\theta_k)
\end{align}
with $m'_f = m'_i + \Lambda$.   This supposes that the wave function peaks sharply enough so that we can evaluate the plane wave amplitude at $\vec p_i = 0$, and do rotations on the nucleon as if it were at rest.  There is only one Wigner function, because we have not yet projected the $\Delta$ onto states with spin quantized along the $z$-axis.

This is the amplitude for producing 
$\ket{\Delta(\vec p_f,m'_f)}$. The whole of the $\Delta$ final state is
\begin{align}
\ket{f} &= \sum_{m'_f} \int \frac{(d^3 p_f)}{2 E_f}
    \ket{\Delta(\vec p_f,m'_f)} (-i)^{m_\gamma}
    e^{i(m_\gamma + m_i) \phi_k - i \vec k \cdot \vec b}
\nn\\       & \times
\widetilde\phi_i(\vec p_f - \vec k)
    \mathcal M^\text{(pw)}_{m'_i \Lambda} \,
        d^{1/2}_{m_i m'_i}(\theta_k)
\end{align}
We get the coordinate wave function of this state using the $\Delta$ field operator and
\begin{align}
\braket{ 0 | \Psi_\mu(x) | \Delta(\vec p_f, m'_f)} =
    u_\mu(\vec p_f,m'_f) \, e^{-i p_f \cdot x}   ,
\end{align}
leaving
\begin{align}
&\braket{ 0 | \Psi_\mu(x) | f } = \sum_{m'_f} 
    \int \frac{ d\phi_k}{2\pi} 
    \frac{ u_\mu(\vec p_f,m'_f) }{ 2 E_f }
    (-i)^{m_\gamma} \nn\\
&\hskip 2 em \times
    e^{i(m_\gamma + m_i) \phi_k - i \vec k \cdot \vec b}    \,
    \phi_i(\vec x) \, \mathcal M^\text{(pw)}_{m'_i \Lambda} \,
        d^{1/2}_{m_i m'_i}(\theta_k)
\end{align}
at time zero.  The peaking of the wave function gives $\vec p_f = \vec k$, and the angles for the outgoing $\Delta$ that we might call $\theta_\Delta,\phi_\Delta$ are the same a $\theta_k,\phi_k$.

The Rarita-Schwinger spin-3/2 spinors are given in terms of spin-1 polarization vectors and spin-1/2 Dirac spinors as
\begin{align}
u_\mu(\vec p,m) = \sum_{\lambda,\mu}
            \left( \begin{array}{c|cc}
            3/2 & 1  &  1/2  \\
            m   & \lambda & \mu
            \end{array}     \right)
    \epsilon_\mu(\hat p,\lambda) u(\vec p,\mu).
\end{align}
Given the structure of the Dirac spinor, one can define a two-spinor with a vector index
\begin{align}
\chi_\mu(\hat p,m) = \sum_{\lambda,\mu}
            \left( \begin{array}{c|cc}
            3/2 & 1  &  1/2  \\
            m   & \lambda & \mu
            \end{array}     \right)
    \epsilon_\mu(\hat p,\lambda) \chi(\hat p,\mu)
\end{align}
where $\chi$ is a two-component spin-1/2 helicity state.  Thence,
\begin{align}
u_\mu(\vec p,m) = \frac{1}{\sqrt{ E_\Delta + M_\Delta }}
    \left(  \begin{array}{c}
         (E_\Delta + M_\Delta) \, \chi_\mu(\hat p,m) \\[1 ex]
         \vec\sigma \cdot \vec p \ \chi_\mu(\hat p,m)
    \end{array}     \right) ,
\end{align}
where the $\vec \sigma$ are the $2\times 2$ Pauli matrices.

Under rotations, $\epsilon_\mu$ and $\chi$ transform simply using the Wigner functions $d^1$ and $d^{1/2}$, respectively, and with Wigner function theorems~\cite{Rose,Edmonds} one can show that
\begin{align}
\chi_\mu(\hat p,m) = R \chi_\mu(\hat z,m)
    = \sum_{m'} e^{-i\phi_k} d^{3/2}_{m m'}(\theta_k) 
        \chi_\mu(\hat z,m') ,
\end{align}
with $R = R(\phi_k,\theta_k,0)$.  This means that the upper components of the Rarita-Schwinger spinor will transform under rotations with the same application of the Wigner functions as for a state at rest.  This means that one can do the same manipulations as for the no-recoil case, and obtain results that look like the nonrelativistic case, Eq.~\eqref{eq:nramplitude}.  This in turn means that although the final $\Delta$ is not a momentum eigenstate, one can express the upper components, and hence the numerical bulk of the state, in terms of momentum eigenstate $u_\mu$ spinors with momenta in the $z$-direction.

However, for the lower components there is additional angular dependence in the $\vec\sigma\cdot\vec p$ term.  In particular,
\begin{align}
\vec\sigma\cdot\hat k = 
\sigma_+ \sin\theta_k e^{-i\phi_k} + 
    \sigma_- \sin\theta_k e^{+i\phi_k} +
        \sigma_z \cos\theta_k   ,
\end{align}
where $\sigma_\pm = \sigma_x \pm i \sigma_y$.  One can still do the azimuthal integral, but the extra $\phi_k$ dependence will bring in Bessel functions with different indices from the main term.

Doing the integral, and projecting onto the $z$-direction spinor $\bar u^\mu(p \hat z,m_f)$ leads to a main term that looks the same as the no-recoil result Eq.~\eqref{eq:nramplitude} plus a correction term which we will have the temerity to write out,
\begin{align}
&\delta \mathcal M = A_0 \, i^{(m_i-m_f)} 
    e^{i(m_i+m_\gamma-m_f)\phi_b} \,
    \frac{M_\Delta-E_\Delta}{2 M_\Delta} \sum_{m'_i}
            \nn\\
&\bigg\{
    \left( \begin{array}{c|cc}
            3/2 & 1  &  1/2  \\
            m_f   & m_f - 1/2 & 1/2
            \end{array}     \right)
    \left( \begin{array}{c|cc}
            3/2 & 1  &  1/2  \\
            m_f - 1   & m_f - 1/2 & -1/2
            \end{array}     \right)     \nn\\[1 ex]
&\hskip 3 em \times    \sin\theta_k  \,
         d^{3/2}_{m_f-1 ,m'_i+\Lambda} (\theta_k)	\,      \nn\\[1.5 ex]
&+   \left( \begin{array}{c|cc}
            3/2 & 1  &  1/2  \\
            m_f   & m_f + 1/2 & -1/2
            \end{array}     \right)
    \left( \begin{array}{c|cc}
            3/2 & 1  &  1/2  \\
            m_f + 1   & m_f + 1/2 & 1/2
            \end{array}     \right)     \nn\\[1 ex]
&\hskip 3 em \times   \sin\theta_k  \,
         d^{3/2}_{m_f+1 ,m'_i+\Lambda} (\theta_k)	\, 
                    \nn\\[1.5 ex]
&+ \  ( \cos\theta_k -1 ) \ 
        d^{3/2}_{m_f ,m'_i+\Lambda} (\theta_k)	\  
\bigg\}                 \nn\\
&\times    \mathcal M^{\text{(pw)}}_{m'_i,\Lambda}  \,
    J_{m_f - m_\gamma - m_i}(\kappa b) \,
        d^{1/2}_{m_i,m'_i}(\theta_k)        \,.
\end{align}
The nominal size of the correction term is
\be
\frac{E_\Delta-M_\Delta}{2M_\Delta} 
    = \frac{k^2}{2 M_\Delta (E_\Delta+M_\Delta)} \approx 1.9\%   \,,
\ee
as already noted in the introduction.

Implementing the corrections, the amplitude with projection onto the final $\Delta$ state with $m_f = 3/2$ is shown in Fig.~\ref{fig:bigamp}(c).  The extra terms have some effect, as seen, but very small.  Corrections for other quantum number situations are visually even smaller.


\section{Rates, cross sections and novel spin effects}		\label{sec:rates}

\begin{figure}[t]
\begin{center}
{\sf (a)} \hfill \,   

\includegraphics[width = 0.85\columnwidth]      {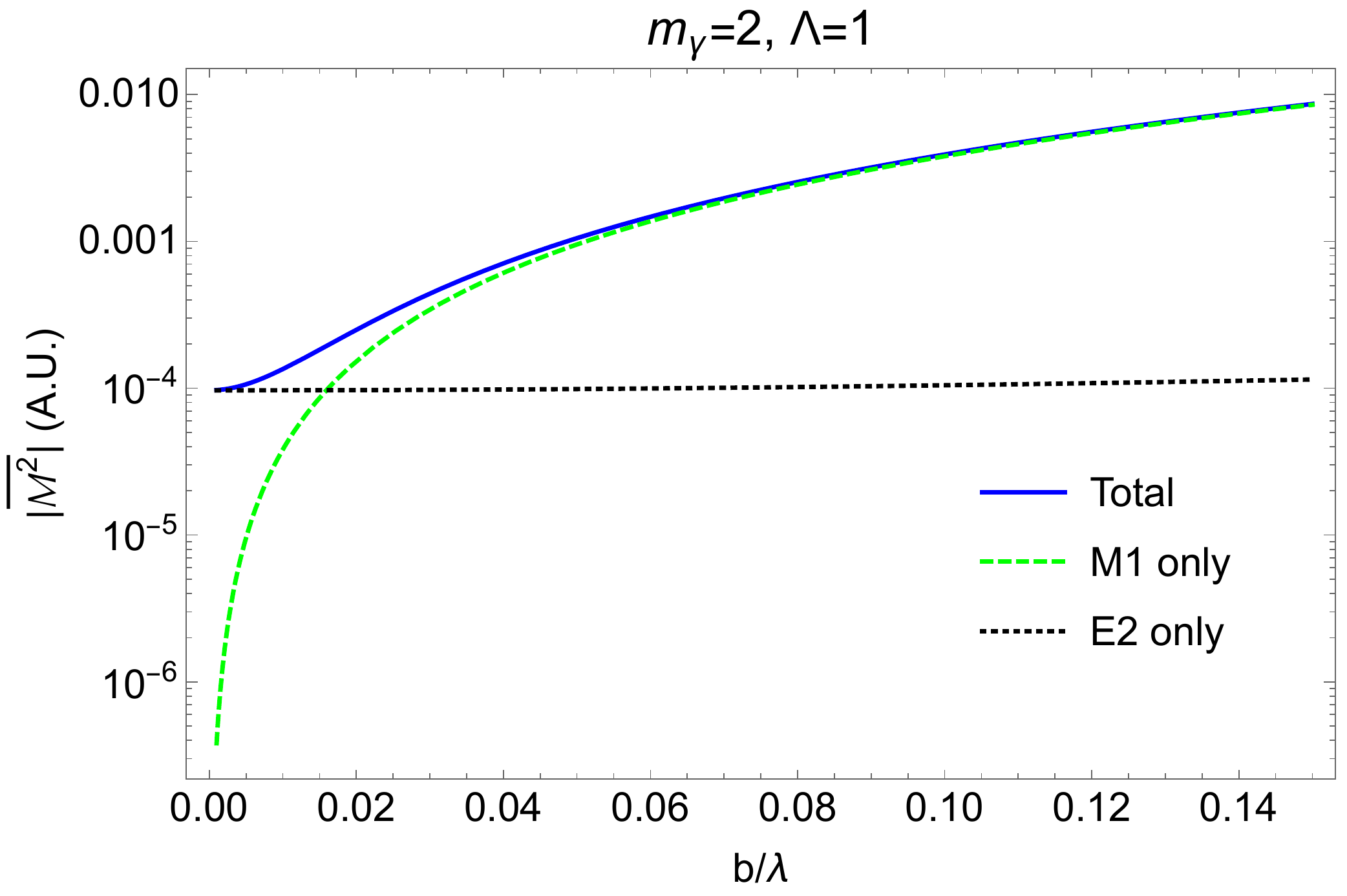}

{\sf (b)} \hfill \,

\includegraphics[width = 0.85\columnwidth]      {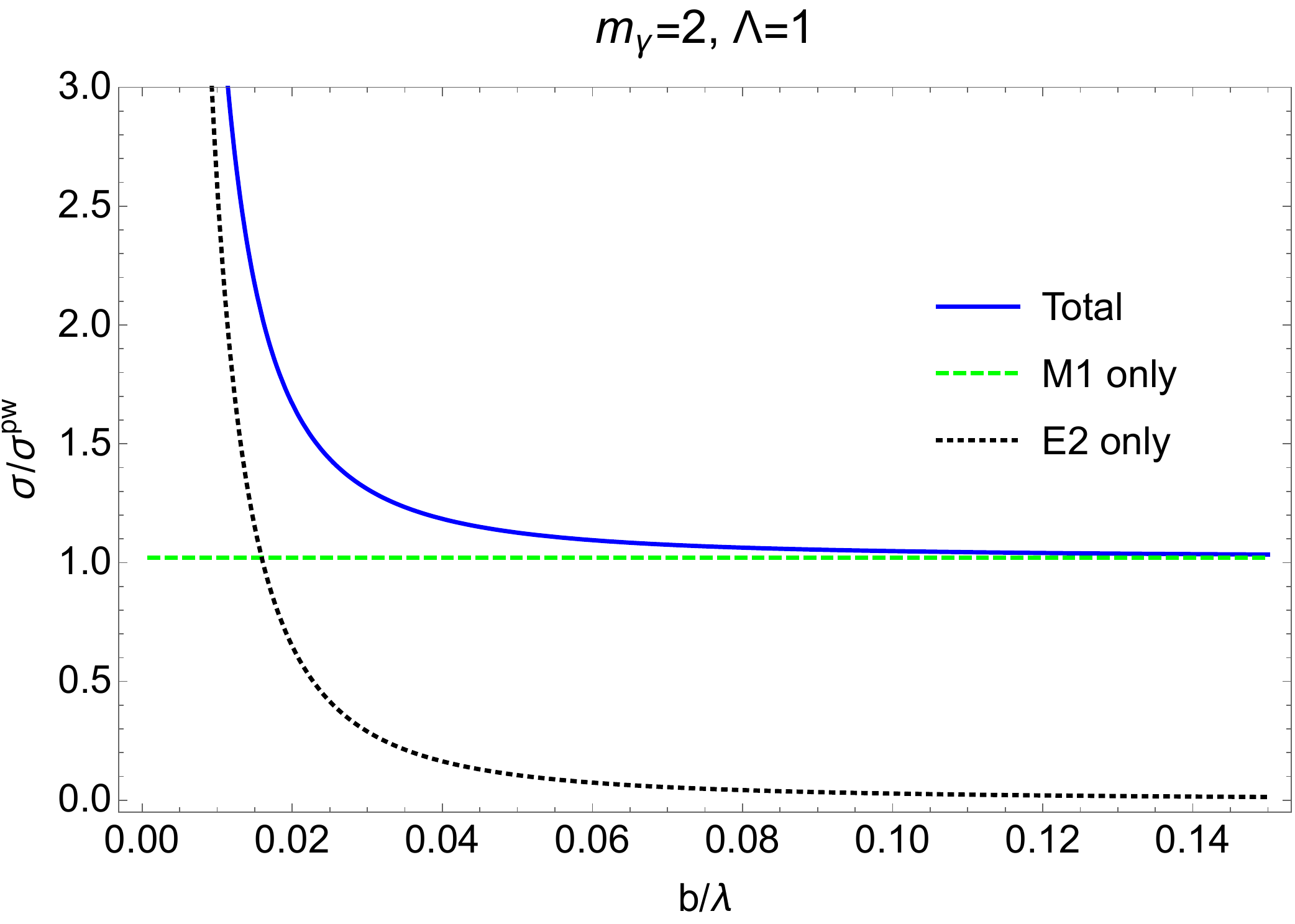}

\caption{(a) Matrix element squared, $|\overline{\mathcal M}|^2$, with summation/averaging performed over baryon helicities. We choose the photon helicity $\Lambda=1$ and total angular momentum projection $m_\gamma=2$. The blue solid line includes both $M1$ and $E2$ transitions, while the green dashed (black dotted) lines show individual contributions from $M1$ ($E2$). (b) Same as (a) divided by the photon energy flux density obtainable from 
Eq.~(\ref{eq:twA}) and normalized to the plane-wave limit $m_\gamma=1$, $\theta_k\to 0$.}
\label{fig:rate}
\end{center}
\end{figure}

Squaring the twisted amplitudes Eq.(\ref{eq:nramplitude}), summing over final helicities of $\Delta$ and averaging over initial helicities of a nucleon, we obtain a quantity $|\overline{\mathcal M}|^2$ that defines a transition rate for photoexcitation. The result is shown in Fig.~\ref{fig:rate}a as a function of proton's transverse position $b$ with respect to the twisted photon's axis. 

It can be seen that the contribution of the electric quadrupole amplitude $E2$ is almost independent of $b$, whereas the magnetic dipole $M1$ contribution is suppressed in the vicinity of the photon vortex center as $b\to 0$. Since this rate is obtained with a position-dependent photon flux, it is instructive to divide the rate by the flux, defining a position-dependent cross section $\sigma(b)$ of photoexcitation shown in Fig.~\ref{fig:rate}b. 

The outcome shows a remarkable feature of the twisted photoexcitation: For $E2$ transitions the rate remains nonzero as $b\to 0$, while the flux turns to zero. It effectively leads to an infinite cross section seen in Fig.~\ref{fig:rate}b at the vortex center.  On the other hand, $M1$ absorption rate is proportional to the flux, same as for the cross-section of plane-wave photoabsorption.  

Next, we compare absorption rates and cross sections for different values of photon helicity $\Lambda$, while keeping fixed the quantity $l_\gamma=m_\gamma-\Lambda$ that corresponds to photon's orbital angular momentum (in a paraxial limit). Namely, we form the following asymmetries:
\be
\label{eq:rateAs}
A^{(\Lambda)}_{\mathcal M^2}=\frac{|\overline{\mathcal M}|^2_{\Lambda=1}-|\overline{\mathcal M}|^2_{\Lambda=-1}}{|\overline{\mathcal M}|^2_{\Lambda=1}+|\overline{\mathcal M}|^2_{\Lambda=-1}} \,
\ee
$i.e.$, the beam spin asymmetry of transition rate and 
\be
\label{eq:xsecAs}
A^{(\Lambda)}_{\sigma}=\frac{\sigma_{\Lambda=1}-\sigma_{\Lambda=-1}}{\sigma_{\Lambda=1}+\sigma_{\Lambda=-1}},\,
\ee
or the beam spin asymmetry of cross sections. Both these asymmetries would be identically zero for plane-wave photons if spatial parity is conserved. They may be nonzero for the twisted photons because they arise due to the differences between twisted photoabsorption with aligned vs. anti-aligned spin ($\Lambda$) with respect to the orbital angular momentum ($l_\gamma$). A similar effect for atomic transitions was predicted theoretically in Ref.\cite{Afanasev_2017}, with $E2$ transitions shown to be responsible for cross section asymmetries.

\begin{figure}[t]
\begin{center}
{\sf (a)} \hfill \,   

\includegraphics[width = 0.85\columnwidth]      {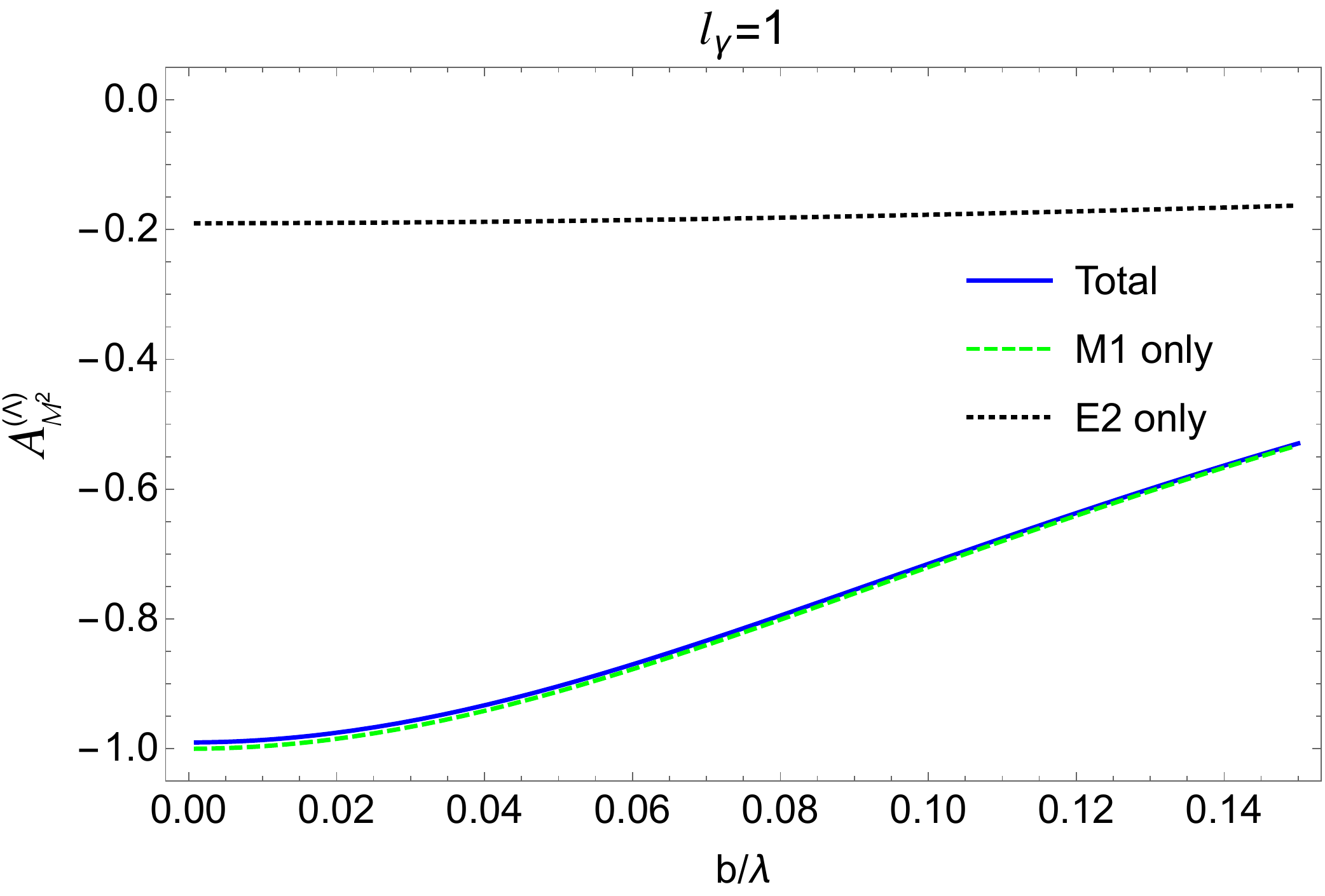}

{\sf (b)} \hfill \,

\includegraphics[width = 0.85\columnwidth]      {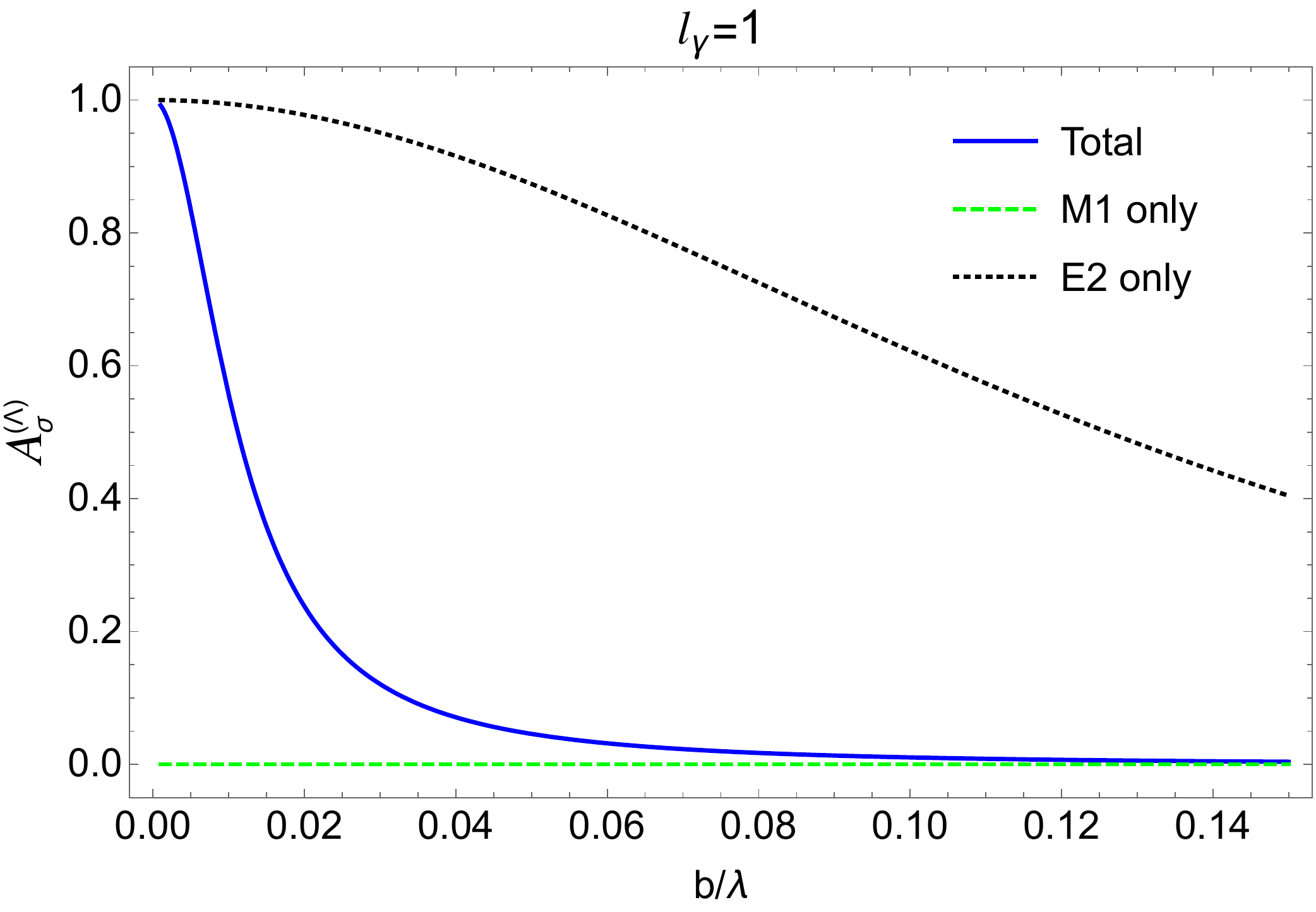}

\caption{Beam spin asymmetry for the transition rate (a) and flux-normalized cross section (b) as defined in Eqs.(\ref{eq:rateAs},\ref{eq:xsecAs}). We took $l_\gamma=1$; notation for the curves is as in Fig.~\ref{fig:rate}.}
\label{fig:asym}
\end{center}
\end{figure}

The results shown in Fig.~\ref{fig:asym} demonstrate that the spin asymmetry of transition rate is dominated by $M1$ transition similarly to spin dependence of the photon flux, but the asymmetry of cross section is dominated by $E2$ contribution near the photon vortex center.

The differences in contributions of $E2$ and $M1$ transitions in the twisted photoabsorption can be better understood from comparing analytic expressions if we perform a Taylor expansion of the cross section near the vortex center in a paraxial limit ($c.f.$ Eqs.(21-24) of Ref.\cite{2018PhRvA..97b3422A}):
\be
\sigma_{(m_\gamma=2,\Lambda=1)}\propto 3G_E^{*2}+G_M^{*2}+\frac{3G_E^{*2}}{(b\pi/\lambda)^2}.
\ee
Comparing the expression above with the plane-wave cross section
\be
\sigma^{pw}\propto 3G_E^{*2}+G_M^{*2},
\ee
it becomes apparent that the twisted cross section has a $1/b^2$ singularity near the vortex center exclusively due to the $E2$ amplitude $G_E^{*}$.

\section{Summary}		\label{sec:end}


We have considered photoproduction of the $\Delta$(1232) baryon by twisted photons.  The angular momentum selection rules for the transitions to definite final states are different from what is possible with plane waves, and leads to additional opportunities.  In particular,  there are transitions to which the main $M1$ component does not contribute at all in the no-recoil or nonrelativistic limit, and even with corrections implemented, contributes only very slightly.  A measurement of these transitions can isolate and measure the smaller $E2$ amplitude, which is a measure of the more complete structure of the baryon states.

Twisted photon experiments like these are clearly for the future, as one needs to learn how to produce energetic twisted photons, with large pitch angles and good control over the location of the photon's vortex axis relative to the target. However, there are good opportunities, and since gamma factories are under consideration,  twisted photon beams should also be considered.


\section*{Acknowledgements}
A.A.~thanks the US Army Research Office Grant W911NF-19-1-0022 for support and C.E.C.~thanks the National Science Foundation (USA) for support under grant PHY-1812326.



\appendix
\section{Larger collection of plots}		\label{sec:appendix}

\begin{figure}[t]
\begin{center}

\includegraphics[width = 0.85 \columnwidth]{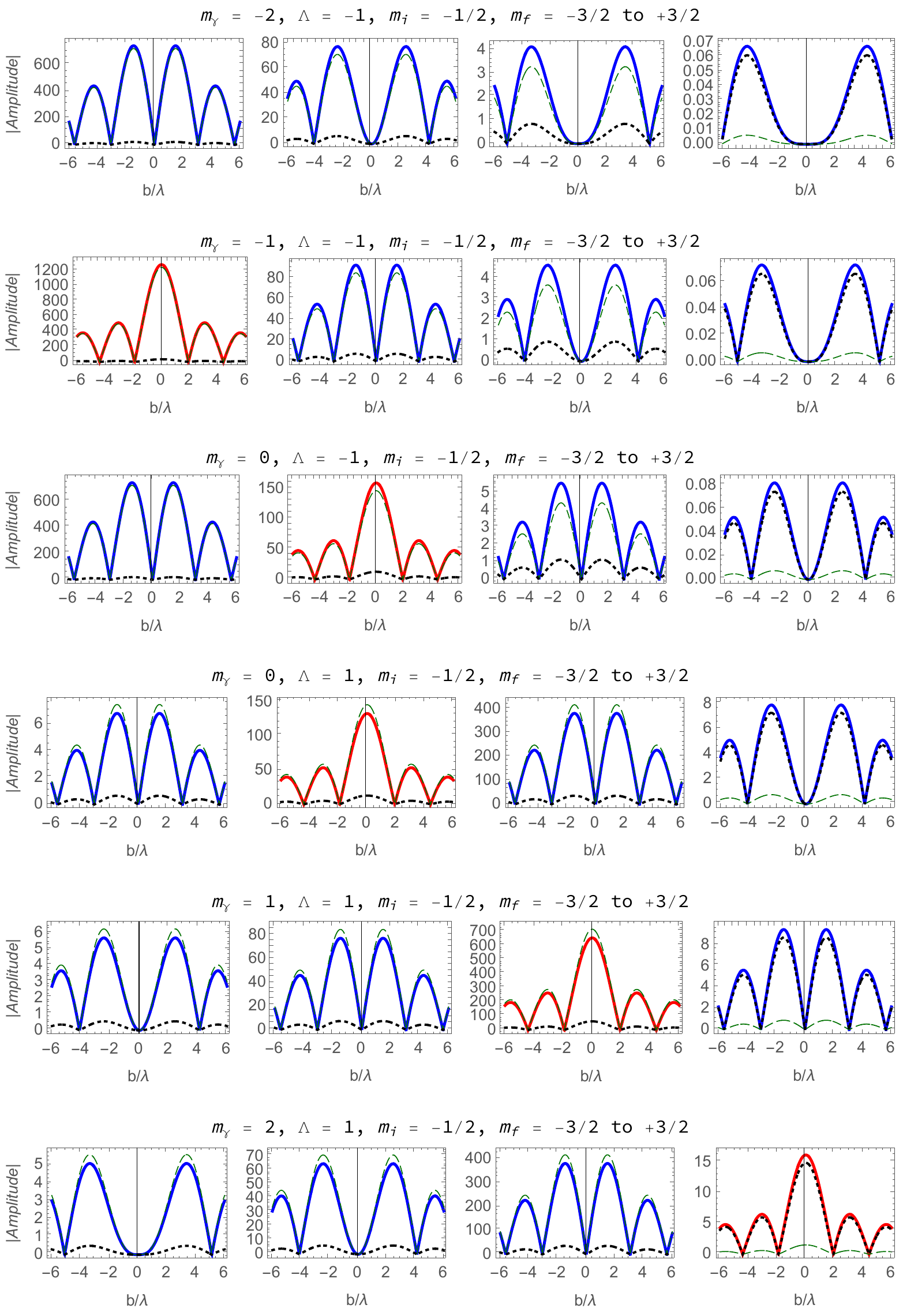}
\caption{Plots of amplitudes for $\gamma+p \to \Delta$ with twisted photons. The protons all have spin projection $m_i = -1/2$.  The $m_\gamma$ and $\Lambda$ for each row are labeled, and the spin projections of the final $\Delta$ run from $-3/2$ to $+3/2$. left to right. The units of the amplitude are arbitrary, but are the same for each plot. The pitch angle for each plot is $\theta_k = 0.2$. The red (for $m_\gamma = \Delta m$) or blue (otherwise) solid curves are the total; the green dashed curves are $G_M^*$ alone and the black dotted curves are $G_E^*$ alone.}
\label{fig:stamps}
\end{center}
\end{figure}

Fig.~\ref{fig:stamps} presents a set of plots for twisted photon + proton $\to \Delta$(1232), with varying choices for the twisted photon total angular momentum $m_\gamma$, photon helicity $\Lambda$, and final $\Delta$ spin projection $m_f$.  The vertical axis shows the amplitude magnitudes, and the horizontal axis shows the impact parameter or the offset between the target location and the photon vortex axis.

All plots have the proton polarized with $m_i = -1/2$.  Plots with $m_i = +1/2$ are identical if one uses parity to change the other quantum numbers appropriately.
Each row has $m_\gamma$ and $\Lambda$, the helicity of the incoming photons in a Fourier decomposition, labeled.  In each row the spin projection of the final $\Delta$ goes from $-3/2$ on the left to $+3/2$ on the right.

These plots include the recoil corrections.  One notices that all plots in the right-hand column show only small contribution from the elsewhere dominant $M1$ amplitude.

\onecolumngrid

\twocolumngrid

\bibliography{deltaphoto}

\begin{thebibliography}{27}%
\makeatletter
\providecommand \@ifxundefined [1]{%
 \@ifx{#1\undefined}
}%
\providecommand \@ifnum [1]{%
 \ifnum #1\expandafter \@firstoftwo
 \else \expandafter \@secondoftwo
 \fi
}%
\providecommand \@ifx [1]{%
 \ifx #1\expandafter \@firstoftwo
 \else \expandafter \@secondoftwo
 \fi
}%
\providecommand \natexlab [1]{#1}%
\providecommand \enquote  [1]{``#1''}%
\providecommand \bibnamefont  [1]{#1}%
\providecommand \bibfnamefont [1]{#1}%
\providecommand \citenamefont [1]{#1}%
\providecommand \href@noop [0]{\@secondoftwo}%
\providecommand \href [0]{\begingroup \@sanitize@url \@href}%
\providecommand \@href[1]{\@@startlink{#1}\@@href}%
\providecommand \@@href[1]{\endgroup#1\@@endlink}%
\providecommand \@sanitize@url [0]{\catcode `\\12\catcode `\$12\catcode
  `\&12\catcode `\#12\catcode `\^12\catcode `\_12\catcode `\%12\relax}%
\providecommand \@@startlink[1]{}%
\providecommand \@@endlink[0]{}%
\providecommand \url  [0]{\begingroup\@sanitize@url \@url }%
\providecommand \@url [1]{\endgroup\@href {#1}{\urlprefix }}%
\providecommand \urlprefix  [0]{URL }%
\providecommand \Eprint [0]{\href }%
\providecommand \doibase [0]{http://dx.doi.org/}%
\providecommand \selectlanguage [0]{\@gobble}%
\providecommand \bibinfo  [0]{\@secondoftwo}%
\providecommand \bibfield  [0]{\@secondoftwo}%
\providecommand \translation [1]{[#1]}%
\providecommand \BibitemOpen [0]{}%
\providecommand \bibitemStop [0]{}%
\providecommand \bibitemNoStop [0]{.\EOS\space}%
\providecommand \EOS [0]{\spacefactor3000\relax}%
\providecommand \BibitemShut  [1]{\csname bibitem#1\endcsname}%
\let\auto@bib@innerbib\@empty
\bibitem [{\citenamefont {Jentschura}\ and\ \citenamefont
  {Serbo}(2011{\natexlab{a}})}]{Jentschura:2011ih}%
  \BibitemOpen
  \bibfield  {author} {\bibinfo {author} {\bibfnamefont {U.}~\bibnamefont
  {Jentschura}}\ and\ \bibinfo {author} {\bibfnamefont {V.}~\bibnamefont
  {Serbo}},\ }\href {\doibase 10.1140/epjc/s10052-011-1571-z} {\bibfield
  {journal} {\bibinfo  {journal} {Eur. Phys. J. C}\ }\textbf {\bibinfo {volume}
  {71}},\ \bibinfo {pages} {1571} (\bibinfo {year} {2011}{\natexlab{a}})},\
  \Eprint {http://arxiv.org/abs/1101.1206} {arXiv:1101.1206 [physics.acc-ph]}
  \BibitemShut {NoStop}%
\bibitem [{\citenamefont {Petrillo}\ \emph {et~al.}(2016)\citenamefont
  {Petrillo}, \citenamefont {Dattoli}, \citenamefont {Drebot},\ and\
  \citenamefont {Nguyen}}]{Petrillo16}%
  \BibitemOpen
  \bibfield  {author} {\bibinfo {author} {\bibfnamefont {V.}~\bibnamefont
  {Petrillo}}, \bibinfo {author} {\bibfnamefont {G.}~\bibnamefont {Dattoli}},
  \bibinfo {author} {\bibfnamefont {I.}~\bibnamefont {Drebot}}, \ and\ \bibinfo
  {author} {\bibfnamefont {F.}~\bibnamefont {Nguyen}},\ }\href {\doibase
  10.1103/PhysRevLett.117.123903} {\bibfield  {journal} {\bibinfo  {journal}
  {Phys. Rev. Lett.}\ }\textbf {\bibinfo {volume} {117}},\ \bibinfo {pages}
  {123903} (\bibinfo {year} {2016})}\BibitemShut {NoStop}%
\bibitem [{\citenamefont {Budker}\ \emph {et~al.}()\citenamefont {Budker},
  \citenamefont {Crespo López-Urrutia}, \citenamefont {Derevianko},
  \citenamefont {Flambaum}, \citenamefont {Krasny}, \citenamefont {Petrenko},
  \citenamefont {Pustelny}, \citenamefont {Surzhykov}, \citenamefont
  {Yerokhin},\ and\ \citenamefont {Zolotorev}}]{Budker20}%
  \BibitemOpen
  \bibfield  {author} {\bibinfo {author} {\bibfnamefont {D.}~\bibnamefont
  {Budker}}, \bibinfo {author} {\bibfnamefont {J.~R.}\ \bibnamefont {Crespo
  López-Urrutia}}, \bibinfo {author} {\bibfnamefont {A.}~\bibnamefont
  {Derevianko}}, \bibinfo {author} {\bibfnamefont {V.~V.}\ \bibnamefont
  {Flambaum}}, \bibinfo {author} {\bibfnamefont {M.~W.}\ \bibnamefont
  {Krasny}}, \bibinfo {author} {\bibfnamefont {A.}~\bibnamefont {Petrenko}},
  \bibinfo {author} {\bibfnamefont {S.}~\bibnamefont {Pustelny}}, \bibinfo
  {author} {\bibfnamefont {A.}~\bibnamefont {Surzhykov}}, \bibinfo {author}
  {\bibfnamefont {V.~A.}\ \bibnamefont {Yerokhin}}, \ and\ \bibinfo {author}
  {\bibfnamefont {M.}~\bibnamefont {Zolotorev}},\ }\href {\doibase
  https://doi.org/10.1002/andp.202000204} {\bibfield  {journal} {\bibinfo
  {journal} {Annalen der Physik}\ }\textbf {\bibinfo {volume} {532}},\ \bibinfo
  {pages} {2000204}}\BibitemShut {NoStop}%
\bibitem [{\citenamefont {Liu}\ \emph {et~al.}(2020)\citenamefont {Liu},
  \citenamefont {Yan}, \citenamefont {Afanasev}, \citenamefont {Benson},
  \citenamefont {Hao}, \citenamefont {Mikhailov}, \citenamefont {Popov},\ and\
  \citenamefont {Wu}}]{liu2020orbital}%
  \BibitemOpen
  \bibfield  {author} {\bibinfo {author} {\bibfnamefont {P.}~\bibnamefont
  {Liu}}, \bibinfo {author} {\bibfnamefont {J.}~\bibnamefont {Yan}}, \bibinfo
  {author} {\bibfnamefont {A.}~\bibnamefont {Afanasev}}, \bibinfo {author}
  {\bibfnamefont {S.~V.}\ \bibnamefont {Benson}}, \bibinfo {author}
  {\bibfnamefont {H.}~\bibnamefont {Hao}}, \bibinfo {author} {\bibfnamefont
  {S.~F.}\ \bibnamefont {Mikhailov}}, \bibinfo {author} {\bibfnamefont {V.~G.}\
  \bibnamefont {Popov}}, \ and\ \bibinfo {author} {\bibfnamefont {Y.~K.}\
  \bibnamefont {Wu}},\ }\href@noop {} {\enquote {\bibinfo {title} {Orbital
  angular momentum beam generation using a free-electron laser oscillator},}\ }
  (\bibinfo {year} {2020}),\ \Eprint {http://arxiv.org/abs/2007.15723}
  {arXiv:2007.15723 [physics.acc-ph]} \BibitemShut {NoStop}%
\bibitem [{\citenamefont {{Drechsler}}\ \emph {et~al.}(2021)\citenamefont
  {{Drechsler}}, \citenamefont {{Wolf}}, \citenamefont {{Schmiegelow}},\ and\
  \citenamefont {{Schmidt-Kaler}}}]{2021arXiv210407095D}%
  \BibitemOpen
  \bibfield  {author} {\bibinfo {author} {\bibfnamefont {M.}~\bibnamefont
  {{Drechsler}}}, \bibinfo {author} {\bibfnamefont {S.}~\bibnamefont {{Wolf}}},
  \bibinfo {author} {\bibfnamefont {C.~T.}\ \bibnamefont {{Schmiegelow}}}, \
  and\ \bibinfo {author} {\bibfnamefont {F.}~\bibnamefont {{Schmidt-Kaler}}},\
  }\href@noop {} {\bibfield  {journal} {\bibinfo  {journal} {arXiv e-prints}\
  ,\ \bibinfo {eid} {arXiv:2104.07095}} (\bibinfo {year} {2021})},\ \Eprint
  {http://arxiv.org/abs/2104.07095} {arXiv:2104.07095 [quant-ph]} \BibitemShut
  {NoStop}%
\bibitem [{\citenamefont {{Schmiegelow}}\ \emph {et~al.}(2016)\citenamefont
  {{Schmiegelow}}, \citenamefont {{Schulz}}, \citenamefont {{Kaufmann}},
  \citenamefont {{Ruster}}, \citenamefont {{Poschinger}},\ and\ \citenamefont
  {{Schmidt-Kaler}}}]{2016NatCo...712998S}%
  \BibitemOpen
  \bibfield  {author} {\bibinfo {author} {\bibfnamefont {C.~T.}\ \bibnamefont
  {{Schmiegelow}}}, \bibinfo {author} {\bibfnamefont {J.}~\bibnamefont
  {{Schulz}}}, \bibinfo {author} {\bibfnamefont {H.}~\bibnamefont
  {{Kaufmann}}}, \bibinfo {author} {\bibfnamefont {T.}~\bibnamefont
  {{Ruster}}}, \bibinfo {author} {\bibfnamefont {U.~G.}\ \bibnamefont
  {{Poschinger}}}, \ and\ \bibinfo {author} {\bibfnamefont {F.}~\bibnamefont
  {{Schmidt-Kaler}}},\ }\href {\doibase 10.1038/ncomms12998} {\bibfield
  {journal} {\bibinfo  {journal} {Nature Communications}\ }\textbf {\bibinfo
  {volume} {7}},\ \bibinfo {eid} {12998} (\bibinfo {year} {2016})}\BibitemShut
  {NoStop}%
\bibitem [{\citenamefont {Afanasev}\ \emph {et~al.}(2018)\citenamefont
  {Afanasev}, \citenamefont {Carlson}, \citenamefont {Schmiegelow},
  \citenamefont {Schulz}, \citenamefont {Schmidt-Kaler},\ and\ \citenamefont
  {Solyanik}}]{Afanasev_2018}%
  \BibitemOpen
  \bibfield  {author} {\bibinfo {author} {\bibfnamefont {A.}~\bibnamefont
  {Afanasev}}, \bibinfo {author} {\bibfnamefont {C.~E.}\ \bibnamefont
  {Carlson}}, \bibinfo {author} {\bibfnamefont {C.~T.}\ \bibnamefont
  {Schmiegelow}}, \bibinfo {author} {\bibfnamefont {J.}~\bibnamefont {Schulz}},
  \bibinfo {author} {\bibfnamefont {F.}~\bibnamefont {Schmidt-Kaler}}, \ and\
  \bibinfo {author} {\bibfnamefont {M.}~\bibnamefont {Solyanik}},\ }\href
  {\doibase 10.1088/1367-2630/aaa63d} {\bibfield  {journal} {\bibinfo
  {journal} {New Journal of Physics}\ }\textbf {\bibinfo {volume} {20}},\
  \bibinfo {pages} {023032} (\bibinfo {year} {2018})}\BibitemShut {NoStop}%
\bibitem [{\citenamefont {Solyanik-Gorgone}\ \emph {et~al.}(2019)\citenamefont
  {Solyanik-Gorgone}, \citenamefont {Afanasev}, \citenamefont {Carlson},
  \citenamefont {Schmiegelow},\ and\ \citenamefont
  {Schmidt-Kaler}}]{Solyanik-Gorgone_2019}%
  \BibitemOpen
  \bibfield  {author} {\bibinfo {author} {\bibfnamefont {M.}~\bibnamefont
  {Solyanik-Gorgone}}, \bibinfo {author} {\bibfnamefont {A.}~\bibnamefont
  {Afanasev}}, \bibinfo {author} {\bibfnamefont {C.~E.}\ \bibnamefont
  {Carlson}}, \bibinfo {author} {\bibfnamefont {C.~T.}\ \bibnamefont
  {Schmiegelow}}, \ and\ \bibinfo {author} {\bibfnamefont {F.}~\bibnamefont
  {Schmidt-Kaler}},\ }\href {\doibase 10.1364/JOSAB.36.000565} {\bibfield
  {journal} {\bibinfo  {journal} {J. Opt. Soc. Am. B}\ }\textbf {\bibinfo
  {volume} {36}},\ \bibinfo {pages} {565} (\bibinfo {year} {2019})}\BibitemShut
  {NoStop}%
\bibitem [{\citenamefont {Pascalutsa}\ \emph {et~al.}(2007)\citenamefont
  {Pascalutsa}, \citenamefont {Vanderhaeghen},\ and\ \citenamefont
  {Yang}}]{Pascalutsa:2006up}%
  \BibitemOpen
  \bibfield  {author} {\bibinfo {author} {\bibfnamefont {V.}~\bibnamefont
  {Pascalutsa}}, \bibinfo {author} {\bibfnamefont {M.}~\bibnamefont
  {Vanderhaeghen}}, \ and\ \bibinfo {author} {\bibfnamefont {S.~N.}\
  \bibnamefont {Yang}},\ }\href {\doibase 10.1016/j.physrep.2006.09.006}
  {\bibfield  {journal} {\bibinfo  {journal} {Phys. Rept.}\ }\textbf {\bibinfo
  {volume} {437}},\ \bibinfo {pages} {125} (\bibinfo {year} {2007})},\ \Eprint
  {http://arxiv.org/abs/hep-ph/0609004} {arXiv:hep-ph/0609004} \BibitemShut
  {NoStop}%
\bibitem [{\citenamefont {Zyla}\ \emph {et~al.}(2020)\citenamefont {Zyla} \emph
  {et~al.}}]{Zyla:2020zbs}%
  \BibitemOpen
  \bibfield  {author} {\bibinfo {author} {\bibfnamefont {P.~A.}\ \bibnamefont
  {Zyla}} \emph {et~al.} (\bibinfo {collaboration} {Particle Data Group}),\
  }\href {\doibase 10.1093/ptep/ptaa104} {\bibfield  {journal} {\bibinfo
  {journal} {PTEP}\ }\textbf {\bibinfo {volume} {2020}},\ \bibinfo {pages}
  {083C01} (\bibinfo {year} {2020})}\BibitemShut {NoStop}%
\bibitem [{\citenamefont {Rynkun}\ \emph {et~al.}(2012)\citenamefont {Rynkun},
  \citenamefont {J{\"o}nsson}, \citenamefont {Gaigalas},\ and\ \citenamefont
  {Fischer}}]{rynkun2012energies}%
  \BibitemOpen
  \bibfield  {author} {\bibinfo {author} {\bibfnamefont {P.}~\bibnamefont
  {Rynkun}}, \bibinfo {author} {\bibfnamefont {P.}~\bibnamefont {J{\"o}nsson}},
  \bibinfo {author} {\bibfnamefont {G.}~\bibnamefont {Gaigalas}}, \ and\
  \bibinfo {author} {\bibfnamefont {C.~F.}\ \bibnamefont {Fischer}},\
  }\href@noop {} {\bibfield  {journal} {\bibinfo  {journal} {Atomic Data and
  Nuclear Data Tables}\ }\textbf {\bibinfo {volume} {98}},\ \bibinfo {pages}
  {481} (\bibinfo {year} {2012})},\ \bibinfo {note} {({T}he specific transition
  is $^2 P_{1/2} \to ^2 \!\!\! D_{3/2}$ (142 nm) on p. 501 of this article.
  Both states have 3 electrons in a $2s 2p^2$ configuration. The atom in
  question is Ne VI, which has 5 electrons, the same as neutral
  Boron.)}\BibitemShut {NoStop}%
\bibitem [{\citenamefont {{Afanasev}}\ \emph {et~al.}(2018)\citenamefont
  {{Afanasev}}, \citenamefont {{Carlson}},\ and\ \citenamefont
  {{Solyanik}}}]{2018PhRvA..97b3422A}%
  \BibitemOpen
  \bibfield  {author} {\bibinfo {author} {\bibfnamefont {A.}~\bibnamefont
  {{Afanasev}}}, \bibinfo {author} {\bibfnamefont {C.~E.}\ \bibnamefont
  {{Carlson}}}, \ and\ \bibinfo {author} {\bibfnamefont {M.}~\bibnamefont
  {{Solyanik}}},\ }\href {\doibase 10.1103/PhysRevA.97.023422} {\bibfield
  {journal} {\bibinfo  {journal} {\pra}\ }\textbf {\bibinfo {volume} {97}},\
  \bibinfo {eid} {023422} (\bibinfo {year} {2018})}\BibitemShut {NoStop}%
\bibitem [{\citenamefont {{Yao}}\ and\ \citenamefont
  {{Padgett}}(2011)}]{2011AdOP....3..161Y}%
  \BibitemOpen
  \bibfield  {author} {\bibinfo {author} {\bibfnamefont {A.~M.}\ \bibnamefont
  {{Yao}}}\ and\ \bibinfo {author} {\bibfnamefont {M.~J.}\ \bibnamefont
  {{Padgett}}},\ }\href {\doibase 10.1364/AOP.3.000161} {\bibfield  {journal}
  {\bibinfo  {journal} {Advances in Optics and Photonics}\ }\textbf {\bibinfo
  {volume} {3}},\ \bibinfo {pages} {161} (\bibinfo {year} {2011})}\BibitemShut
  {NoStop}%
\bibitem [{\citenamefont {Bliokh}\ and\ \citenamefont
  {Nori}(2015)}]{Bliokh:2015doa}%
  \BibitemOpen
  \bibfield  {author} {\bibinfo {author} {\bibfnamefont {K.~Y.}\ \bibnamefont
  {Bliokh}}\ and\ \bibinfo {author} {\bibfnamefont {F.}~\bibnamefont {Nori}},\
  }\href {\doibase 10.1016/j.physrep.2015.06.003} {\bibfield  {journal}
  {\bibinfo  {journal} {Phys. Rept.}\ }\textbf {\bibinfo {volume} {592}},\
  \bibinfo {pages} {1} (\bibinfo {year} {2015})},\ \Eprint
  {http://arxiv.org/abs/1504.03113} {arXiv:1504.03113 [physics.optics]}
  \BibitemShut {NoStop}%
\bibitem [{\citenamefont {Jentschura}\ and\ \citenamefont
  {Serbo}(2011{\natexlab{b}})}]{Jentschura:2010ap}%
  \BibitemOpen
  \bibfield  {author} {\bibinfo {author} {\bibfnamefont {U.}~\bibnamefont
  {Jentschura}}\ and\ \bibinfo {author} {\bibfnamefont {V.}~\bibnamefont
  {Serbo}},\ }\href {\doibase 10.1103/PhysRevLett.106.013001} {\bibfield
  {journal} {\bibinfo  {journal} {Phys. Rev. Lett.}\ }\textbf {\bibinfo
  {volume} {106}},\ \bibinfo {pages} {013001} (\bibinfo {year}
  {2011}{\natexlab{b}})}\BibitemShut {NoStop}%
\bibitem [{\citenamefont {Afanasev}\ \emph {et~al.}(2013)\citenamefont
  {Afanasev}, \citenamefont {Carlson},\ and\ \citenamefont
  {Mukherjee}}]{Afanasev:2013kaa}%
  \BibitemOpen
  \bibfield  {author} {\bibinfo {author} {\bibfnamefont {A.}~\bibnamefont
  {Afanasev}}, \bibinfo {author} {\bibfnamefont {C.~E.}\ \bibnamefont
  {Carlson}}, \ and\ \bibinfo {author} {\bibfnamefont {A.}~\bibnamefont
  {Mukherjee}},\ }\href {\doibase 10.1103/PhysRevA.88.033841} {\bibfield
  {journal} {\bibinfo  {journal} {Phys. Rev. A}\ }\textbf {\bibinfo {volume}
  {88}},\ \bibinfo {pages} {033841} (\bibinfo {year} {2013})}\BibitemShut
  {NoStop}%
\bibitem [{\citenamefont {{Quinteiro}}\ \emph {et~al.}(2019)\citenamefont
  {{Quinteiro}}, \citenamefont {{Schmiegelow}}, \citenamefont {{Reiter}},\ and\
  \citenamefont {{Kuhn}}}]{2019PhRvA..99b3845Q}%
  \BibitemOpen
  \bibfield  {author} {\bibinfo {author} {\bibfnamefont {G.~F.}\ \bibnamefont
  {{Quinteiro}}}, \bibinfo {author} {\bibfnamefont {C.~T.}\ \bibnamefont
  {{Schmiegelow}}}, \bibinfo {author} {\bibfnamefont {D.~E.}\ \bibnamefont
  {{Reiter}}}, \ and\ \bibinfo {author} {\bibfnamefont {T.}~\bibnamefont
  {{Kuhn}}},\ }\href {\doibase 10.1103/PhysRevA.99.023845} {\bibfield
  {journal} {\bibinfo  {journal} {\pra}\ }\textbf {\bibinfo {volume} {99}},\
  \bibinfo {eid} {023845} (\bibinfo {year} {2019})}\BibitemShut {NoStop}%
\bibitem [{\citenamefont {Scholz-Marggraf}\ \emph {et~al.}(2014)\citenamefont
  {Scholz-Marggraf}, \citenamefont {Fritzsche}, \citenamefont {Serbo},
  \citenamefont {Afanasev},\ and\ \citenamefont
  {Surzhykov}}]{scholz2014absorption}%
  \BibitemOpen
  \bibfield  {author} {\bibinfo {author} {\bibfnamefont {H.~M.}\ \bibnamefont
  {Scholz-Marggraf}}, \bibinfo {author} {\bibfnamefont {S.}~\bibnamefont
  {Fritzsche}}, \bibinfo {author} {\bibfnamefont {V.~G.}\ \bibnamefont
  {Serbo}}, \bibinfo {author} {\bibfnamefont {A.}~\bibnamefont {Afanasev}}, \
  and\ \bibinfo {author} {\bibfnamefont {A.}~\bibnamefont {Surzhykov}},\ }\href
  {\doibase 10.1103/PhysRevA.90.013425} {\bibfield  {journal} {\bibinfo
  {journal} {Phys. Rev. A}\ }\textbf {\bibinfo {volume} {90}},\ \bibinfo
  {pages} {013425} (\bibinfo {year} {2014})}\BibitemShut {NoStop}%
\bibitem [{\citenamefont {Wick}(1962)}]{Wick:1962zz}%
  \BibitemOpen
  \bibfield  {author} {\bibinfo {author} {\bibfnamefont {G.~C.}\ \bibnamefont
  {Wick}},\ }\href {\doibase 10.1016/0003-4916(62)90059-3} {\bibfield
  {journal} {\bibinfo  {journal} {Annals Phys.}\ }\textbf {\bibinfo {volume}
  {18}},\ \bibinfo {pages} {65} (\bibinfo {year} {1962})}\BibitemShut {NoStop}%
\bibitem [{\citenamefont {Jones}\ and\ \citenamefont
  {Scadron}(1973)}]{Jones:1972ky}%
  \BibitemOpen
  \bibfield  {author} {\bibinfo {author} {\bibfnamefont {H.~F.}\ \bibnamefont
  {Jones}}\ and\ \bibinfo {author} {\bibfnamefont {M.~D.}\ \bibnamefont
  {Scadron}},\ }\href {\doibase 10.1016/0003-4916(73)90476-4} {\bibfield
  {journal} {\bibinfo  {journal} {Annals Phys.}\ }\textbf {\bibinfo {volume}
  {81}},\ \bibinfo {pages} {1} (\bibinfo {year} {1973})}\BibitemShut {NoStop}%
\bibitem [{jon()}]{jonesscadronnote}%
  \BibitemOpen
  \href@noop {} {}\bibinfo {note} {Eq.~\eqref{eq:pwamplitudes} was obtained
  using Ref.~\cite{Pascalutsa:2006up} and agrees with~\cite{Jones:1972ky} if
  one identifies $\mathcal H(0)$ here with $- i e \Gamma$ there and also
  inserts the isospin factor.}\BibitemShut {Stop}%
\bibitem [{\citenamefont {{Afanasev}}\ \emph {et~al.}(2014)\citenamefont
  {{Afanasev}}, \citenamefont {{Carlson}},\ and\ \citenamefont
  {{Mukherjee}}}]{2014JOSAB..31.2721A}%
  \BibitemOpen
  \bibfield  {author} {\bibinfo {author} {\bibfnamefont {A.}~\bibnamefont
  {{Afanasev}}}, \bibinfo {author} {\bibfnamefont {C.~E.}\ \bibnamefont
  {{Carlson}}}, \ and\ \bibinfo {author} {\bibfnamefont {A.}~\bibnamefont
  {{Mukherjee}}},\ }\href {\doibase 10.1364/JOSAB.31.002721} {\bibfield
  {journal} {\bibinfo  {journal} {Journal of the Optical Society of America B
  Optical Physics}\ }\textbf {\bibinfo {volume} {31}},\ \bibinfo {pages} {2721}
  (\bibinfo {year} {2014})}\BibitemShut {NoStop}%
\bibitem [{\citenamefont {Taylor}()}]{Taylor}%
  \BibitemOpen
  \bibfield  {author} {\bibinfo {author} {\bibfnamefont {J.~R.}\ \bibnamefont
  {Taylor}},\ }\href@noop {} {\emph {\bibinfo {title} {{Scattering Theory: The
  Quantum Theory of Nonrelativistic Collisions}}}}\ (\bibinfo  {publisher}
  {Dover, Mineola (2006)})\ \bibinfo {note} {\!\!, (originally Wiley, New York
  (1972) and also Krieger, Malabar, Florida (1983)); particularly see chapters
  2 and 3}\BibitemShut {NoStop}%
\bibitem [{\citenamefont {Peskin}\ and\ \citenamefont
  {Schroeder}(1995)}]{Peskin:1995ev}%
  \BibitemOpen
  \bibfield  {author} {\bibinfo {author} {\bibfnamefont {M.~E.}\ \bibnamefont
  {Peskin}}\ and\ \bibinfo {author} {\bibfnamefont {D.~V.}\ \bibnamefont
  {Schroeder}},\ }\href@noop {} {\emph {\bibinfo {title} {{An Introduction to
  quantum field theory}}}}\ (\bibinfo  {publisher} {Addison-Wesley},\ \bibinfo
  {address} {Reading, USA},\ \bibinfo {year} {1995})\ \bibinfo {note} {\!\!,
  particularly section 4.5}\BibitemShut {NoStop}%
\bibitem [{\citenamefont {Rose}(1957)}]{Rose}%
  \BibitemOpen
  \bibfield  {author} {\bibinfo {author} {\bibfnamefont {M.~E.}\ \bibnamefont
  {Rose}},\ }\href@noop {} {\emph {\bibinfo {title} {Elementary Theory of
  Angular Momentum}}}\ (\bibinfo  {publisher} {John Wiley and Sons},\ \bibinfo
  {address} {New York},\ \bibinfo {year} {1957})\BibitemShut {NoStop}%
\bibitem [{\citenamefont {Edmonds}(1957)}]{Edmonds}%
  \BibitemOpen
  \bibfield  {author} {\bibinfo {author} {\bibfnamefont {A.~R.}\ \bibnamefont
  {Edmonds}},\ }\href@noop {} {\emph {\bibinfo {title} {Angular Momentum in
  Quantum Mechanics}}}\ (\bibinfo  {publisher} {Princeton University Press},\
  \bibinfo {address} {Princeton, NJ},\ \bibinfo {year} {1957})\BibitemShut
  {NoStop}%
\bibitem [{\citenamefont {Afanasev}\ \emph {et~al.}(2017)\citenamefont
  {Afanasev}, \citenamefont {Carlson},\ and\ \citenamefont
  {Solyanik}}]{Afanasev_2017}%
  \BibitemOpen
  \bibfield  {author} {\bibinfo {author} {\bibfnamefont {A.}~\bibnamefont
  {Afanasev}}, \bibinfo {author} {\bibfnamefont {C.~E.}\ \bibnamefont
  {Carlson}}, \ and\ \bibinfo {author} {\bibfnamefont {M.}~\bibnamefont
  {Solyanik}},\ }\href {\doibase 10.1088/2040-8986/aa82c3} {\bibfield
  {journal} {\bibinfo  {journal} {Journal of Optics}\ }\textbf {\bibinfo
  {volume} {19}},\ \bibinfo {pages} {105401} (\bibinfo {year}
  {2017})}\BibitemShut {NoStop}%
\end{thebibliography}%

\end{document}